\documentclass[aps, prb, showpacs, preprint, amsmath, letterpaper]{revtex4-1}

\usepackage{graphics}
\usepackage{graphicx}
\usepackage{amssymb}
\usepackage{bm}

\begin{document}
\title{Enhanced spin triplet superconductivity due to Kondo destabilization}

\author{Sheng Ran$^{1,2,3*}$, Hyunsoo Kim$^{1*}$, I-Lin Liu$^{1,2,3*}$, Shanta Saha$^{1,2}$, Ian Hayes$^{1}$, Tristin Metz$^{1}$, Yun Suk Eo$^{1}$, Johnpierre Paglione$^{1,2}$, Nicholas P. Butch$^{1,2}$}

\affiliation{$^1$ Center for Nanophysics and Advanced Materials, Department of Physics, University of Maryland, College Park, MD 20742, USA
\\$^2$ NIST Center for Neutron Research, National Institute of Standards and Technology, Gaithersburg, MD 20899, USA
\\$^3$ Department of Materials Science and Engineering, University of Maryland, College Park, MD 20742, USA
\\$^\ast$ Equally contributed
}

\date{\today}

\begin{abstract}
In a Kondo lattice system, suppression of effective Kondo coupling leads to the breakdown of the heavy-electron metal and a change in the electronic structure~\cite{Si2001,Coleman2001,Paul2007,Yang2017}. Spin triplet superconductivity in the Kondo lattice UTe$_2$~\cite{Ran2019} appears to be associated with spin fluctuations originating from incipient ferromagnetic order. Here we show clear evidence of twofold enhancement of spin-triplet superconductivity near the pressure-driven suppression of the Kondo coherence, implying that superconductivity is strengthened by the affiliated growth of both spin and charge fluctuations. The coherent Kondo state discontinuously transitions to ferromagnetic order at higher pressures. Application of  magnetic field tunes the system back across a first-order phase boundary. Straddling this phase boundary, we find another example of reentrant superconductivity in UTe$_2$~\cite{Ran2019NP}. In addition to spin fluctuations~\cite{Tokunaga2019,Sundar2019} associated with ferromagnetism,  our results show that a Kondo-driven Fermi surface instability may be playing a role in stabilizing spin triplet superconductivity.

\end{abstract}
\maketitle
While proximity to antiferromagnetism is believed to be a key ingredient for unconventional superconductivity (SC), ferromagnetism (FM) is generally antagonistic and incompatible with superconductivity. In a very few cases~\cite{Saxena2000,Aoki2001,Huy2007}, where FM and SC coexist and are carried by the same electrons, magnetic fluctuations tend to induce triplet pairing, which is a natural candidate for topological SC~\cite{Sato2017}. Understanding the mechanisms that helps to stabilize triplet SC is therefore important both at the fundamental quantum mechanics level as well as for potential application for quantum computation.  

The recently discovered heavy fermion superconductor UTe$_2$~\cite{Ran2019,Aoki2019}, as a paramagnetic end member of the ferromagnetic superconductor series, provides a new platform to study the interaction between FM and triplet SC. The triplet pairing in UTe$_2$ is clearly manifested by striking experimental results: a remarkably large and anisotropic upper critical field; temperature independent nuclear magnetic resonance (NMR) Knight shift; two independent reentrant superconducting phases existing in extremely high magnetic fields~\cite{Ran2019NP,Knebel2019}; point node gap structure demonstrated by thermal conductivity, penetration depth~\cite{Metz2019} and specific heat measurements~\cite{Ran2019,Aoki2019}. Scanning tunneling microscopy further reveals signatures of chiral in-gap states predicted to exist on the boundary of a topological superconductor~\cite{Jiao2019}. 

Unlike the ferromagnetic superconductors that share some common features with UTe$_2$~\cite{Saxena2000,Aoki2001,Huy2007}, UTe$_2$ does not order magnetically prior to the onset of SC~\cite{Ran2019,Hutanu2019}. Instead, scaling analysis shows that it is close to FM quantum criticality~\cite{Ran2019}. Strong, nearly critical fluctuations have been revealed by NMR~\cite{Tokunaga2019} and muon spin relaxation ($\mu$SR) measurements~\cite{Sundar2019}. Therefore, a quantum phase transition into a magnetic phase is likely to be revealed by tuning the system with pressure. 

Here we report two fold enhancement of spin triplet SC in UTe$_2$~\cite{Ran2019} under pressure. This occurs as the Kondo coherence is continuously suppressed, towards an apparent quantum critical point associated with a Fermi surface instability. However, this trend with pressure terminates at a first order transition, and at higher pressures, magnetic order emerges. This phase boundary can be crossed again by applying magnetic field which increases hybridization, and SC reenters. This shows that both spin fluctuations and electronic structure changes conspire to strengthen SC.

\begin{figure}
\includegraphics[angle=0,width=180mm]{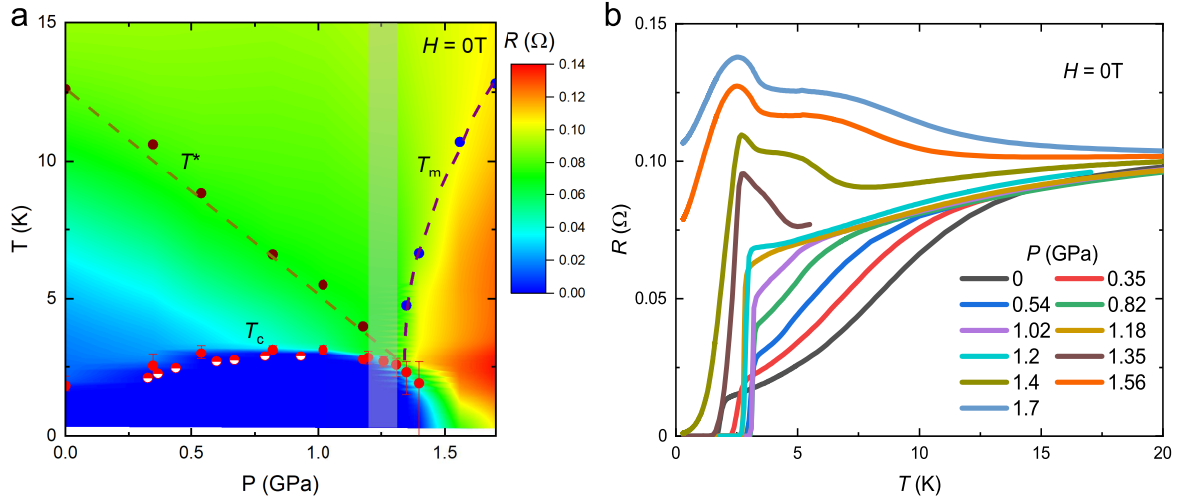}
\caption{Phase diagram of UTe$_2$ under pressure in zero magnetic field. (a) Color contour plot of the resistivity data as a function of both temperature and pressures in zero magnetic field, and the resulting phase diagram. Solid red dots represent $T_c$ of superconductivity determined from resistance measurements. Error bars of $T_c$ are defined by the onset and offset of superconducting transition. The half open red dots represent $T_c$ of superconductivity determined from magnetization measurements. Brown dots represents the kinks in $R(T)$ data in the low pressure range. Blue dots represent the local minimum in $R(T)$ data in the high pressure range. The grey region indicates the critical pressure region of finite width. (b) The temperature dependence of resistivity data in zero magnetic field for selected pressure values. The low-temperature resistivity exhibits a clear evolution in slope, from positive to negative, as pressure increases.}
\label{H0}
\end{figure}

Fig.\ref{H0}a summarizes the resistivity data as a function of both temperature and pressure in zero magnetic field. Below 1.31 GPa, the transition temperature of SC, $T_c$, forms a clear dome feature under pressure peaked at 1~GPa, where $T_c$ is doubled, compared to the ambient pressure value, reaching 3.2~K. The bulk nature of the SC is confirmed by magnetization data under pressure up to 0.93~GPa, measured down to 1.8~K (supplement). 

The enhancement of $T_c$ is accompanied by a systematic change in the low-temperature normal state resistance value (Fig.\ref{H0}b). At ambient pressure, the resistivity in the normal state continuously decreases and shows a slope change. The temperature of this slope change $T^*$, is very sensitive to the current direction. In this study, the current flows in the (011) plane, and the slope change appears at 13~K at ambient pressure. There is no magnetic or structural phase transition in this temperature range, as demonstrated by specific heat and neutron diffraction measurements~\cite{Ran2019,Hutanu2019}, so this reduction in electrical resistivity is due to the establishment of Kondo coherence. As pressure increases, $T^*$ is monotonically suppressed from 13~K to about 5~K for 1.02~GPa, and at higher pressures the signature for Kondo coherence is no longer visible. Resistivity above $T^*$ mostly originates from single ion Kondo behavior. Magneto-resistance is then due to the polarization of single-ion Kondo isolated magnetic atoms, and is governed by one energy scale. Therefore, $R(H)$ curves should scale with a temperature dependent effective field $H*$~\cite{Batlogg1987,Brison1989}. As shown in the supplement, good scaling is achieved for temperatures above $T^*$, and it starts to deviate at low temperatures. It is also clear that the energy scale is suppressed by pressure, e.g., for 0.45~GPa, the scaling is achieved above 10~K, while for 1.18~GPa, the scaling works from temperatures just above $T_c$. Suppression of collective hybridization of local moments leads to a breakdown of the heavy Fermi liquid, and a jump in size of Fermi surface~\cite{Si2001,Coleman2001,Paul2007,Yang2017}. This Kondo destabilization produces both spin and charge fluctuations that could separately or collectively promote attractive interactions to enhance SC~\cite{Park2008}.

\begin{figure}
\includegraphics[angle=0,width=180mm]{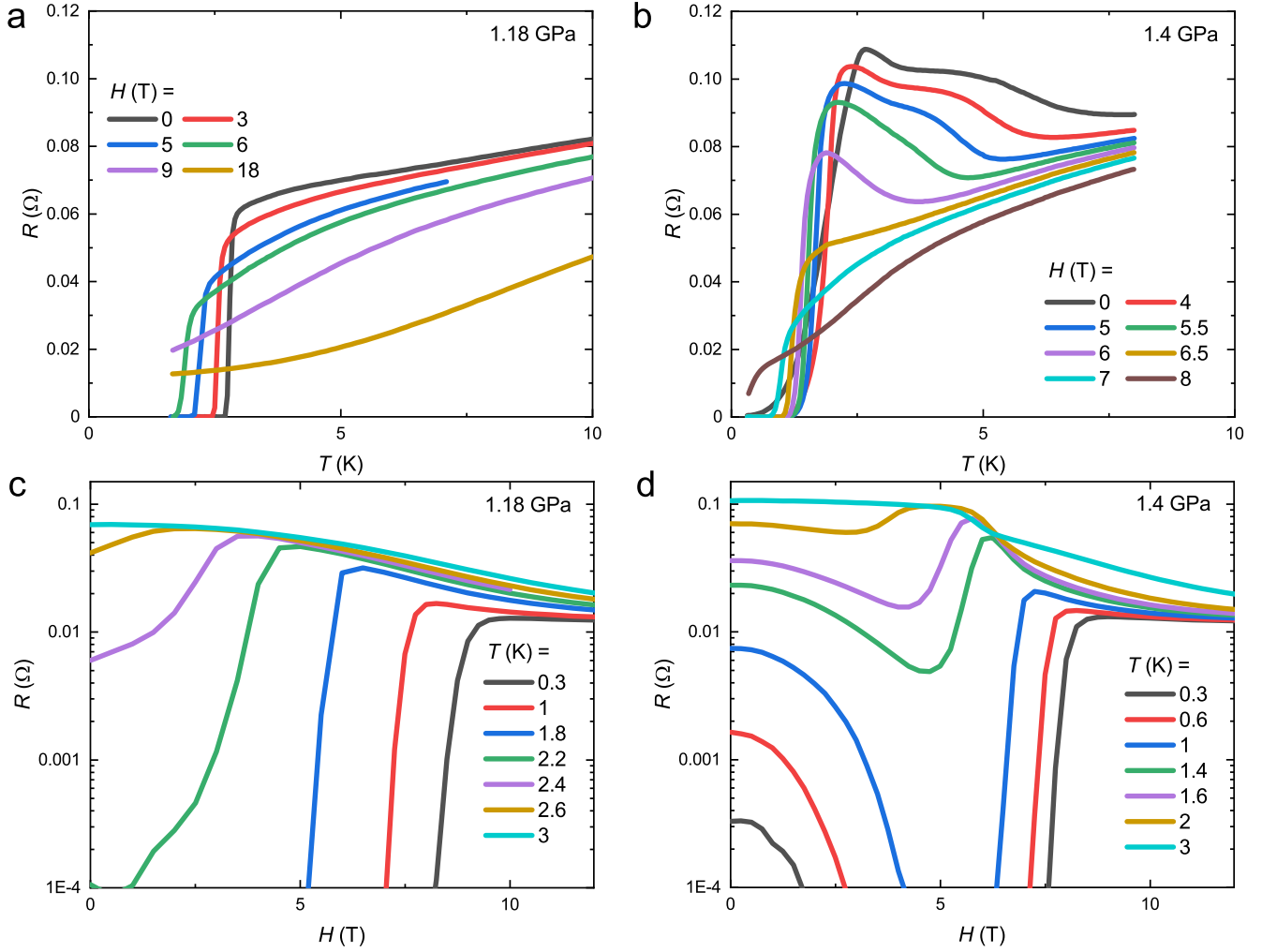}
\caption{Magnetic field as a tuning parameter. (a)-(b) Resistance data as a function of temperature for different magnetic fields, for 1.18 and 1.4~GPa. Negative normal-state magnetoresistance and sharp upper critical fields are evident. (c)-(d) Resistance data as a function of magnetic field for different temperatures, for 1.18 and 1.4~GPa. Reentrant SC is readily apparent in the low temperature magnetoresistance.}
\label{RTH}
\end{figure}

As the pressure further increases, both the normal state and superconducting properties change dramatically. The normal state resistivity increases upon cooling with two successive local minima indicating phase transitions. The temperature of the local minimum at higher temperature $T_M$ increases with pressure, while that of the lower temperature minimum is roughly pressure independent. $T_M$ is highly sensitive to the magnetic field, e.g., suppressed from 7~K to 4~K by 6~T, and disappears in higher magnetic field for 1.4~GPa, indicating its magnetic nature (Fig.\ref{RTH}b). The coincidence of the onset of increasing resistance and magnetic order (Fig. 1a) show that the two phenomena stem from the suppression of the Kondo hybridization that follows $T^*$. Neither $T^*$ nor $T_M$ appear to track to zero temperature. Extrapolations of the pressure dependence of $T^*$, $T_M$, and $T_c$ meet in the critical pressure region, suggesting that a first-order transition occurs when these phenomena have a common finite energy scale.


In an interesting twist, magnetism is suppressed by applied magnetic field, resistivity decreases, and SC is induced, yielding another example of reentrant SC in UTe$_2$. This is most apparent at 1.4~GPa. At this pressure, although there is a large drop in the resistivity at low temperatures, a zero resistance state is not achieved (Fig~\ref{RTH}). This is a signature of partial volume SC, which is stabilized by local strains on the high-pressure side of the first order phase transition. As magnetic field is increased, the resistivity finally drops to zero. This reentrant SC is stable between fields of 2~T and 8~T, and appears to be related to the sharply suppressed magnetic order. Similar reentrance of SC in the magnetic field is also observed for 1.35~GPa, but only at 1.6~K, not the zero temperature limit.

\begin{figure}
\includegraphics[angle=0,width=160mm]{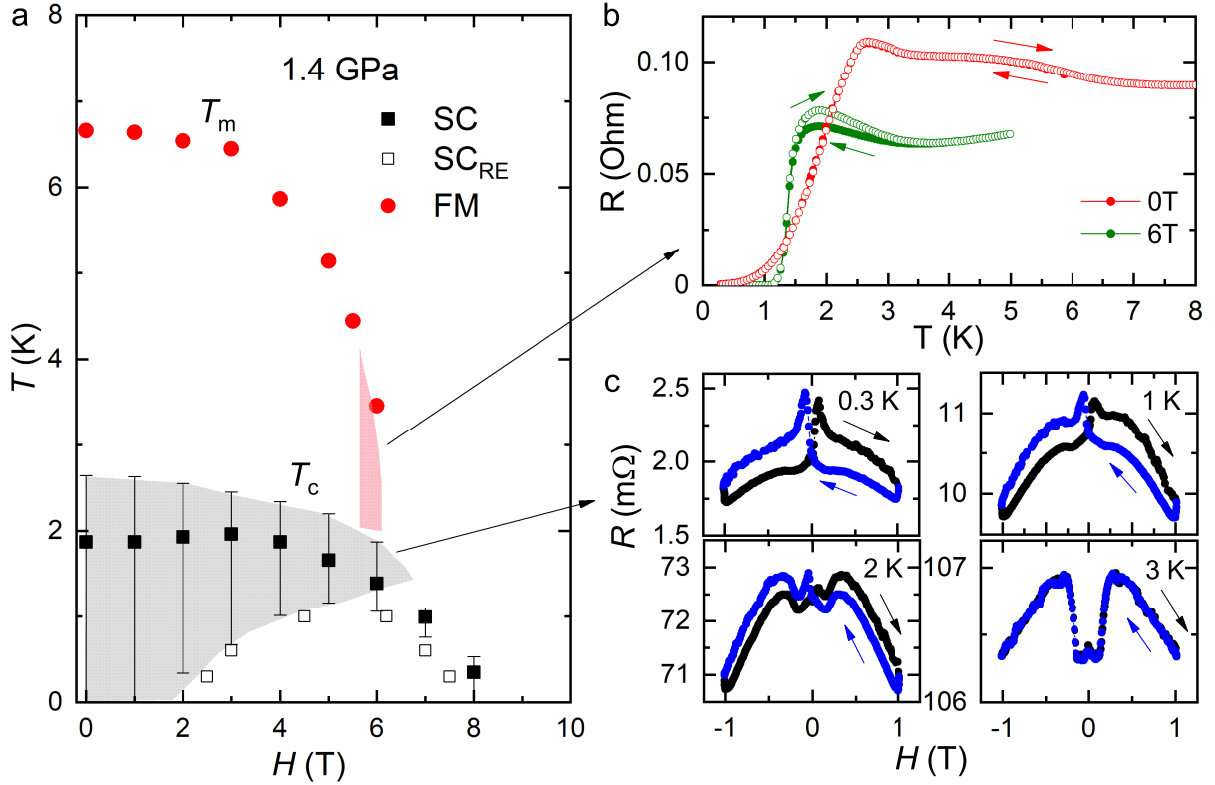}
\caption{Hysteresis in temperature and magnetic field for 1.4~GPa. (a) Magnetic and superconducting phase boundaries at 1.4~GPa. In the pink region hysteresis is observed in the temperature dependent resistance data. In the grey region, superconductivity coexists with magnetism, and hysteresis is observed in the field dependent resistance data. Error bars of $T_c$ are defined by the onset and offset of superconducting transition. (b) Resistance data as a function of temperature for $H$ = 0 and 6~T, showing clear hysteresis in temperature when the magnetic phase is suppressed to low enough temperature. (c) Resistance data as a function of magnetic field from -1 to 1~T for 0.3, 1, 2 and 3~K, showing hysteresis in magnetic field in the region where superconductivity coexists with magnetism.}
\label{Hysteresis}
\end{figure}

In the region of partial volume SC (grey region in Fig.~\ref{Hysteresis}a), we observe fairly large hysteresis in magnetic field dependence of $R$ data (Fig.~\ref{Hysteresis}c). Below 2~K, The resistivity increases quickly upon up sweep in the very low field range, leading to a larger value than that upon down sweep. Above the grey region, the hysteresis disappears. Such hysteresis is typically associated with FM domain motion, indicating the magnetism under high pressure is FM. On the other hand, the hysteresis observed here is only seen at temperatures below the sudden drop of resistivity, suggesting the role of an additional mechanism. Due to the first order phase transition separating SC and FM as a function of pressure, both phases can coexist heterogeneously. The relative volume fractions are different upon up and down sweep of magnetic field, leading to the hysteresis. The first order nature of $T_M$ is more obvious when it is suppressed to lower temperatures by applied field. As shown in Fig.~\ref{Hysteresis}b, in 6~T, $R(T)$ also shows well defined hysteresis in temperature. Similar hysteresis is observed at lower pressure in zero field (supplement). As FM quantum phase transitions are discontinuous in clean metallic systems~\cite{Brando2016} yet can still act as a source of strong order parameter fluctuations~\cite{Schmakat2015}, this critical pressure-field region emerges as a likely source of the strong spin fluctuations observed in UTe$_{2}$ at ambient pressure.

Two other reentrant superconducting phases have been already observed in UTe$_{2}$ under high magnetic field~\cite{Ran2019NP,Knebel2019}, at ambient pressure, which are likely induced by ferromagnetic fluctuations and decreased dimensionality. The reentrant superconducting phase observed under pressure is quite different. First, the magnetic field scale is much smaller here. In addition, in this case reentrant SC exists on both side of the magnetism boundary, while in case of the field induced SC at ambient pressure, SC only exists in the field polarized state~\cite{Ran2019NP}. These differences indicate the reentrance of SC is probably due to different mechanism. The dome like feature and the suppression of two energy scales in the vicinity of optimal SC: Kondo coherence temperature and magnetism, indicate SC is closely related with to magnetic and charge fluctuations.



The fact that applied magnetic field can tune the relative strength of Kondo hybridization and long range magnetic order provides a natural explanation for the reentrance of SC. Global phase stability of a Kondo lattice system is governed by two quantities, the ratio between the Kondo interaction and the RKKY interaction, and the degree of quantum fluctuations of the local-moment magnetism~\cite{Si2006,Si2010}. Destabilizing long range magnetic order with applied field tunes the system back into the critical regime, although in UTe$_2$ the stability of SC weakens away from $P_c$.

\begin{figure}
\includegraphics[angle=0,width=150mm]{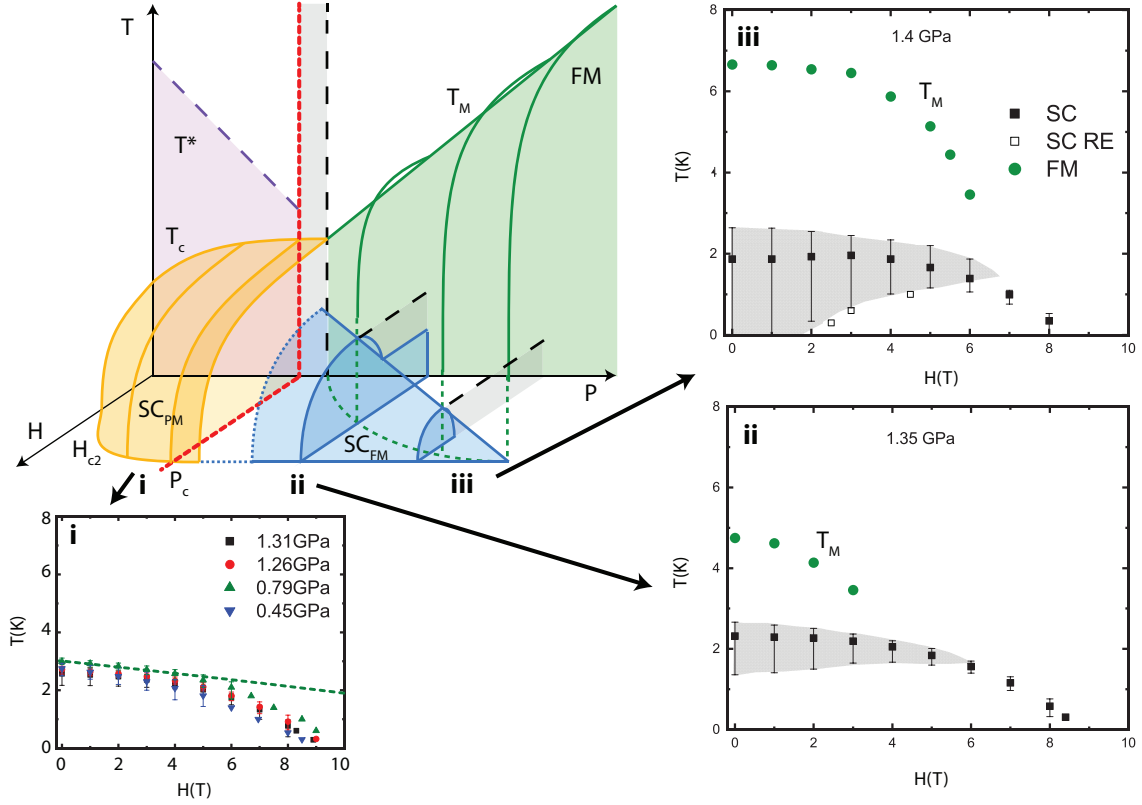}
\caption{Schematic phase diagram of UTe$_2$, emphasizing the opposing roles of pressure $P$ and magnetic field $H$ as tuning parameters. In the low pressure region, paramagnetic superconductivity SC$_{PM}$ (in yellow) coexists below $T^*$ with a Kondo coherent electronic structure (purple). On the high pressure side of the critical pressure $P_c$, ferromagnetic order (green) is suppressed by field and reentrant superconductivity SC$_{FM}$ is observed at low temperature (blue). Coexistence of FM and SC is observed at these pressures.  Constant pressure slices are shown for low pressure (i), 1.35~GPa (ii) and 1.4~GPa (iii). In (i), the $T$ and $H$ limits of superconductivity are rather pressure-insensitive. In (ii) and (iii), the FM/SC coexistence regions are marked in gray, and the relationship between optimal SC and suppression of FM are clearly seen. Error bars of $T_c$ are defined by the onset and offset of superconducting transition.}
\label{Schematic}
\end{figure}

The pressure dependence of UTe$_2$ is qualitatively different from that of the ferromagnetic superconductors~\cite{Aoki2019a}, in which cases SC coexists with FM. For UGe$_2$ and URhGe, SC exclusively exists inside the FM region, while for UCoGe, SC exists on both sides of the FM boundary~\cite{Aoki2019a}. In all these cases, FM fluctuations are believed to be responsible for the triplet pairing~\cite{Mineev2017a}. In general, the role of electronic instabilities in the ferromagnetic superconductors remains an open question; even in the case of UGe$_2$, where changes in magnetic order coincide with apparent Fermi surface changes, SC is not a ground state on the paramagnetic side~\cite{Aoki2019a}. However, in pressure-tuned UTe$_2$, SC and FM exist on two opposite sides of the phase boundary (Fig.~\ref{Schematic}), and our results indicate that destabilizing Kondo coupling is responsible for enhancing the SC on the paramagnetic side. This insight may help to better understand the FM superconductors and further reveals a new paradigm for enhancing spin triplet SC.

\section{methods}
Single crystals of UTe$_{2}$ were synthesized by the chemical vapor transport method using iodine as the transport agent. A non-magnetic piston-cylinder pressure cell was used for electrical transport measurements under pressure up to 1.7 GPa, with Daphne oil as the pressure medium. Transport measurements were performed in a Quantum Design Physical Property Measurement System (PPMS), and in Oxford $^3$He system. Magnetic susceptibility measurements under hydrostatic pressure were performed in a Quantum Design Magnetic Property Measurement System (MPMS) using a BeCu piston-cylinder clamp cell with Daphne oil as pressure medium. In both cases,pressure produced on the single crystal sample at low temperatures were calibrated by measuring the superconducting transition temperature of lead placed in the cell. The known pressure dependencies of the superconducting transition temperature of Pb~\cite{Smith1967} were used for this purpose.

\section{Data Availability}
All other data that support the plots within this paper and other findings of this study are available from the corresponding author upon reasonable request.

\section{addendum}
We acknowledge S. L. Bud'ko and P. C. Canfield for providing pressure medium. S. Ran is grateful for inspiring discussions with Y. F. Yang and Y. Wang. We acknowledge W. T. Fuhrman for assistance during sample synthesis. We also acknowledge H. Hodovanets for helpful assistance during our experiments. The Research at the University of Maryland was supported by the US National Science Foundation (NSF) Division of Materials Research Award No. DMR-1610349, the US Department of Energy (DOE) Award No. DE-SC-0019154 (experimental investigations), and the Gordon and Betty Moore Foundation’s EPiQS Initiative through Grant No. GBMF4419 (materials synthesis).

Identification of commercial equipment does not imply recommendation or endorsement by NIST.



\bibliography{UTe2pressure}

\begin{thebibliography}{27}%
\makeatletter
\providecommand \@ifxundefined [1]{%
 \@ifx{#1\undefined}
}%
\providecommand \@ifnum [1]{%
 \ifnum #1\expandafter \@firstoftwo
 \else \expandafter \@secondoftwo
 \fi
}%
\providecommand \@ifx [1]{%
 \ifx #1\expandafter \@firstoftwo
 \else \expandafter \@secondoftwo
 \fi
}%
\providecommand \natexlab [1]{#1}%
\providecommand \enquote  [1]{``#1''}%
\providecommand \bibnamefont  [1]{#1}%
\providecommand \bibfnamefont [1]{#1}%
\providecommand \citenamefont [1]{#1}%
\providecommand \href@noop [0]{\@secondoftwo}%
\providecommand \href [0]{\begingroup \@sanitize@url \@href}%
\providecommand \@href[1]{\@@startlink{#1}\@@href}%
\providecommand \@@href[1]{\endgroup#1\@@endlink}%
\providecommand \@sanitize@url [0]{\catcode `\\12\catcode `\$12\catcode
  `\&12\catcode `\#12\catcode `\^12\catcode `\_12\catcode `\%12\relax}%
\providecommand \@@startlink[1]{}%
\providecommand \@@endlink[0]{}%
\providecommand \url  [0]{\begingroup\@sanitize@url \@url }%
\providecommand \@url [1]{\endgroup\@href {#1}{\urlprefix }}%
\providecommand \urlprefix  [0]{URL }%
\providecommand \Eprint [0]{\href }%
\providecommand \doibase [0]{http://dx.doi.org/}%
\providecommand \selectlanguage [0]{\@gobble}%
\providecommand \bibinfo  [0]{\@secondoftwo}%
\providecommand \bibfield  [0]{\@secondoftwo}%
\providecommand \translation [1]{[#1]}%
\providecommand \BibitemOpen [0]{}%
\providecommand \bibitemStop [0]{}%
\providecommand \bibitemNoStop [0]{.\EOS\space}%
\providecommand \EOS [0]{\spacefactor3000\relax}%
\providecommand \BibitemShut  [1]{\csname bibitem#1\endcsname}%
\let\auto@bib@innerbib\@empty
\bibitem [{\citenamefont {Si}\ \emph {et~al.}(2001)\citenamefont {Si},
  \citenamefont {Rabello}, \citenamefont {Ingersent},\ and\ \citenamefont
  {Smith}}]{Si2001}%
  \BibitemOpen
  \bibfield  {author} {\bibinfo {author} {\bibfnamefont {Q.}~\bibnamefont
  {Si}}, \bibinfo {author} {\bibfnamefont {S.}~\bibnamefont {Rabello}},
  \bibinfo {author} {\bibfnamefont {K.}~\bibnamefont {Ingersent}}, \ and\
  \bibinfo {author} {\bibfnamefont {J.~L.}\ \bibnamefont {Smith}},\ }\href
  {https://doi.org/10.1038/35101507} {\bibfield  {journal} {\bibinfo  {journal}
  {Nature}\ }\textbf {\bibinfo {volume} {413}},\ \bibinfo {pages} {804}
  (\bibinfo {year} {2001})}\BibitemShut {NoStop}%
\bibitem [{\citenamefont {Coleman}\ \emph {et~al.}(2001)\citenamefont
  {Coleman}, \citenamefont {Pépin}, \citenamefont {Si},\ and\ \citenamefont
  {Ramazashvili}}]{Coleman2001}%
  \BibitemOpen
  \bibfield  {author} {\bibinfo {author} {\bibfnamefont {P.}~\bibnamefont
  {Coleman}}, \bibinfo {author} {\bibfnamefont {C.}~\bibnamefont {Pépin}},
  \bibinfo {author} {\bibfnamefont {Q.}~\bibnamefont {Si}}, \ and\ \bibinfo
  {author} {\bibfnamefont {R.}~\bibnamefont {Ramazashvili}},\ }\href
  {http://dx.doi.org/10.1088/0953-8984/13/35/202} {\bibfield  {journal}
  {\bibinfo  {journal} {Journal of Physics: Condensed Matter}\ }\textbf
  {\bibinfo {volume} {13}},\ \bibinfo {pages} {R723} (\bibinfo {year}
  {2001})}\BibitemShut {NoStop}%
\bibitem [{\citenamefont {Paul}\ \emph {et~al.}(2007)\citenamefont {Paul},
  \citenamefont {Pépin},\ and\ \citenamefont {Norman}}]{Paul2007}%
  \BibitemOpen
  \bibfield  {author} {\bibinfo {author} {\bibfnamefont {I.}~\bibnamefont
  {Paul}}, \bibinfo {author} {\bibfnamefont {C.}~\bibnamefont {Pépin}}, \ and\
  \bibinfo {author} {\bibfnamefont {M.~R.}\ \bibnamefont {Norman}},\ }\href
  {https://link.aps.org/doi/10.1103/PhysRevLett.98.026402} {\bibfield
  {journal} {\bibinfo  {journal} {PRL}\ }\textbf {\bibinfo {volume} {98}},\
  \bibinfo {pages} {026402} (\bibinfo {year} {2007})}\BibitemShut {NoStop}%
\bibitem [{\citenamefont {Yang}\ \emph {et~al.}(2017)\citenamefont {Yang},
  \citenamefont {Pines},\ and\ \citenamefont {Lonzarich}}]{Yang2017}%
  \BibitemOpen
  \bibfield  {author} {\bibinfo {author} {\bibfnamefont {Y.-f.}\ \bibnamefont
  {Yang}}, \bibinfo {author} {\bibfnamefont {D.}~\bibnamefont {Pines}}, \ and\
  \bibinfo {author} {\bibfnamefont {G.}~\bibnamefont {Lonzarich}},\ }\href
  {\doibase 10.1073/pnas.1703172114} {\bibfield  {journal} {\bibinfo  {journal}
  {Proceedings of the National Academy of Sciences}\ }\textbf {\bibinfo
  {volume} {114}},\ \bibinfo {pages} {6250} (\bibinfo {year} {2017})},\ \Eprint
  {http://arxiv.org/abs/https://www.pnas.org/content/114/24/6250.full.pdf}
  {https://www.pnas.org/content/114/24/6250.full.pdf} \BibitemShut {NoStop}%
\bibitem [{\citenamefont {Ran}\ \emph {et~al.}(2019{\natexlab{a}})\citenamefont
  {Ran}, \citenamefont {Eckberg}, \citenamefont {Ding}, \citenamefont
  {Furukawa}, \citenamefont {Metz}, \citenamefont {Saha}, \citenamefont {Liu},
  \citenamefont {Zic}, \citenamefont {Kim}, \citenamefont {Paglione},\ and\
  \citenamefont {Butch}}]{Ran2019}%
  \BibitemOpen
  \bibfield  {author} {\bibinfo {author} {\bibfnamefont {S.}~\bibnamefont
  {Ran}}, \bibinfo {author} {\bibfnamefont {C.}~\bibnamefont {Eckberg}},
  \bibinfo {author} {\bibfnamefont {Q.-P.}\ \bibnamefont {Ding}}, \bibinfo
  {author} {\bibfnamefont {Y.}~\bibnamefont {Furukawa}}, \bibinfo {author}
  {\bibfnamefont {T.}~\bibnamefont {Metz}}, \bibinfo {author} {\bibfnamefont
  {S.~R.}\ \bibnamefont {Saha}}, \bibinfo {author} {\bibfnamefont {I.-L.}\
  \bibnamefont {Liu}}, \bibinfo {author} {\bibfnamefont {M.}~\bibnamefont
  {Zic}}, \bibinfo {author} {\bibfnamefont {H.}~\bibnamefont {Kim}}, \bibinfo
  {author} {\bibfnamefont {J.}~\bibnamefont {Paglione}}, \ and\ \bibinfo
  {author} {\bibfnamefont {N.~P.}\ \bibnamefont {Butch}},\ }\href {\doibase
  10.1126/science.aav8645} {\bibfield  {journal} {\bibinfo  {journal}
  {Science}\ }\textbf {\bibinfo {volume} {365}},\ \bibinfo {pages} {684}
  (\bibinfo {year} {2019}{\natexlab{a}})},\ \Eprint
  {http://arxiv.org/abs/https://science.sciencemag.org/content/365/6454/684.full.pdf}
  {https://science.sciencemag.org/content/365/6454/684.full.pdf} \BibitemShut
  {NoStop}%
\bibitem [{\citenamefont {Ran}\ \emph {et~al.}(2019{\natexlab{b}})\citenamefont
  {Ran}, \citenamefont {Liu}, \citenamefont {Eo}, \citenamefont {Campbell},
  \citenamefont {Neves}, \citenamefont {Fuhrman}, \citenamefont {Saha},
  \citenamefont {Eckberg}, \citenamefont {Kim}, \citenamefont {Paglione},
  \citenamefont {Graf}, \citenamefont {Singleton},\ and\ \citenamefont
  {Butch}}]{Ran2019NP}%
  \BibitemOpen
  \bibfield  {author} {\bibinfo {author} {\bibfnamefont {S.}~\bibnamefont
  {Ran}}, \bibinfo {author} {\bibfnamefont {I.-L.}\ \bibnamefont {Liu}},
  \bibinfo {author} {\bibfnamefont {Y.~S.}\ \bibnamefont {Eo}}, \bibinfo
  {author} {\bibfnamefont {D.~J.}\ \bibnamefont {Campbell}}, \bibinfo {author}
  {\bibfnamefont {P.}~\bibnamefont {Neves}}, \bibinfo {author} {\bibfnamefont
  {W.~T.}\ \bibnamefont {Fuhrman}}, \bibinfo {author} {\bibfnamefont {S.~R.}\
  \bibnamefont {Saha}}, \bibinfo {author} {\bibfnamefont {C.}~\bibnamefont
  {Eckberg}}, \bibinfo {author} {\bibfnamefont {H.}~\bibnamefont {Kim}},
  \bibinfo {author} {\bibfnamefont {J.}~\bibnamefont {Paglione}}, \bibinfo
  {author} {\bibfnamefont {D.}~\bibnamefont {Graf}}, \bibinfo {author}
  {\bibfnamefont {J.}~\bibnamefont {Singleton}}, \ and\ \bibinfo {author}
  {\bibfnamefont {N.~P.}\ \bibnamefont {Butch}},\ }\href@noop {} {\bibfield
  {journal} {\bibinfo  {journal} {Nature Physics, accepted}\ } (\bibinfo {year}
  {2019}{\natexlab{b}})}\BibitemShut {NoStop}%
\bibitem [{\citenamefont {Tokunaga}\ \emph {et~al.}(2019)\citenamefont
  {Tokunaga}, \citenamefont {Sakai}, \citenamefont {Kambe}, \citenamefont
  {Hattori}, \citenamefont {Higa}, \citenamefont {Nakamine}, \citenamefont
  {Kitagawa}, \citenamefont {Ishida}, \citenamefont {Nakamura}, \citenamefont
  {Shimizu}, \citenamefont {Homma}, \citenamefont {Li}, \citenamefont {Honda},\
  and\ \citenamefont {Aoki}}]{Tokunaga2019}%
  \BibitemOpen
  \bibfield  {author} {\bibinfo {author} {\bibfnamefont {Y.}~\bibnamefont
  {Tokunaga}}, \bibinfo {author} {\bibfnamefont {H.}~\bibnamefont {Sakai}},
  \bibinfo {author} {\bibfnamefont {S.}~\bibnamefont {Kambe}}, \bibinfo
  {author} {\bibfnamefont {T.}~\bibnamefont {Hattori}}, \bibinfo {author}
  {\bibfnamefont {N.}~\bibnamefont {Higa}}, \bibinfo {author} {\bibfnamefont
  {G.}~\bibnamefont {Nakamine}}, \bibinfo {author} {\bibfnamefont
  {S.}~\bibnamefont {Kitagawa}}, \bibinfo {author} {\bibfnamefont
  {K.}~\bibnamefont {Ishida}}, \bibinfo {author} {\bibfnamefont
  {A.}~\bibnamefont {Nakamura}}, \bibinfo {author} {\bibfnamefont
  {Y.}~\bibnamefont {Shimizu}}, \bibinfo {author} {\bibfnamefont
  {Y.}~\bibnamefont {Homma}}, \bibinfo {author} {\bibfnamefont
  {D.}~\bibnamefont {Li}}, \bibinfo {author} {\bibfnamefont {F.}~\bibnamefont
  {Honda}}, \ and\ \bibinfo {author} {\bibfnamefont {D.}~\bibnamefont {Aoki}},\
  }\href@noop {} {\bibfield  {journal} {\bibinfo  {journal} {arXiv}\ }
  (\bibinfo {year} {2019})}\BibitemShut {NoStop}%
\bibitem [{\citenamefont {Sundar}\ \emph {et~al.}(2019)\citenamefont {Sundar},
  \citenamefont {Gheidi}, \citenamefont {Akintola}, \citenamefont {Cote},
  \citenamefont {Dunsiger}, \citenamefont {Ran}, \citenamefont {Butch},
  \citenamefont {Saha}, \citenamefont {Paglione},\ and\ \citenamefont
  {Sonier}}]{Sundar2019}%
  \BibitemOpen
  \bibfield  {author} {\bibinfo {author} {\bibfnamefont {S.}~\bibnamefont
  {Sundar}}, \bibinfo {author} {\bibfnamefont {S.}~\bibnamefont {Gheidi}},
  \bibinfo {author} {\bibfnamefont {K.}~\bibnamefont {Akintola}}, \bibinfo
  {author} {\bibfnamefont {A.~M.}\ \bibnamefont {Cote}}, \bibinfo {author}
  {\bibfnamefont {S.~R.}\ \bibnamefont {Dunsiger}}, \bibinfo {author}
  {\bibfnamefont {S.}~\bibnamefont {Ran}}, \bibinfo {author} {\bibfnamefont
  {N.~P.}\ \bibnamefont {Butch}}, \bibinfo {author} {\bibfnamefont {S.~R.}\
  \bibnamefont {Saha}}, \bibinfo {author} {\bibfnamefont {J.}~\bibnamefont
  {Paglione}}, \ and\ \bibinfo {author} {\bibfnamefont {J.~E.}\ \bibnamefont
  {Sonier}},\ }\href@noop {} {\bibfield  {journal} {\bibinfo  {journal}
  {arXiv}\ } (\bibinfo {year} {2019})}\BibitemShut {NoStop}%
\bibitem [{\citenamefont {Saxena}\ \emph {et~al.}(2000)\citenamefont {Saxena},
  \citenamefont {Agarwal}, \citenamefont {Ahilan}, \citenamefont {Grosche},
  \citenamefont {Haselwimmer}, \citenamefont {Steiner}, \citenamefont {Pugh},
  \citenamefont {Walker}, \citenamefont {Julian}, \citenamefont {Monthoux},
  \citenamefont {Lonzarich}, \citenamefont {Huxley}, \citenamefont {Sheikin},
  \citenamefont {Braithwaite},\ and\ \citenamefont {Flouquet}}]{Saxena2000}%
  \BibitemOpen
  \bibfield  {author} {\bibinfo {author} {\bibfnamefont {S.~S.}\ \bibnamefont
  {Saxena}}, \bibinfo {author} {\bibfnamefont {P.}~\bibnamefont {Agarwal}},
  \bibinfo {author} {\bibfnamefont {K.}~\bibnamefont {Ahilan}}, \bibinfo
  {author} {\bibfnamefont {F.~M.}\ \bibnamefont {Grosche}}, \bibinfo {author}
  {\bibfnamefont {R.~K.~W.}\ \bibnamefont {Haselwimmer}}, \bibinfo {author}
  {\bibfnamefont {M.~J.}\ \bibnamefont {Steiner}}, \bibinfo {author}
  {\bibfnamefont {E.}~\bibnamefont {Pugh}}, \bibinfo {author} {\bibfnamefont
  {I.~R.}\ \bibnamefont {Walker}}, \bibinfo {author} {\bibfnamefont {S.~R.}\
  \bibnamefont {Julian}}, \bibinfo {author} {\bibfnamefont {P.}~\bibnamefont
  {Monthoux}}, \bibinfo {author} {\bibfnamefont {G.~G.}\ \bibnamefont
  {Lonzarich}}, \bibinfo {author} {\bibfnamefont {A.}~\bibnamefont {Huxley}},
  \bibinfo {author} {\bibfnamefont {I.}~\bibnamefont {Sheikin}}, \bibinfo
  {author} {\bibfnamefont {D.}~\bibnamefont {Braithwaite}}, \ and\ \bibinfo
  {author} {\bibfnamefont {J.}~\bibnamefont {Flouquet}},\ }\href
  {http://dx.doi.org/10.1038/35020500} {\bibfield  {journal} {\bibinfo
  {journal} {Nature}\ }\textbf {\bibinfo {volume} {406}},\ \bibinfo {pages}
  {587} (\bibinfo {year} {2000})}\BibitemShut {NoStop}%
\bibitem [{\citenamefont {Aoki}\ \emph {et~al.}(2001)\citenamefont {Aoki},
  \citenamefont {Huxley}, \citenamefont {Ressouche}, \citenamefont
  {Braithwaite}, \citenamefont {Flouquet}, \citenamefont {Brison},
  \citenamefont {Lhotel},\ and\ \citenamefont {Paulsen}}]{Aoki2001}%
  \BibitemOpen
  \bibfield  {author} {\bibinfo {author} {\bibfnamefont {D.}~\bibnamefont
  {Aoki}}, \bibinfo {author} {\bibfnamefont {A.}~\bibnamefont {Huxley}},
  \bibinfo {author} {\bibfnamefont {E.}~\bibnamefont {Ressouche}}, \bibinfo
  {author} {\bibfnamefont {D.}~\bibnamefont {Braithwaite}}, \bibinfo {author}
  {\bibfnamefont {J.}~\bibnamefont {Flouquet}}, \bibinfo {author}
  {\bibfnamefont {J.-P.}\ \bibnamefont {Brison}}, \bibinfo {author}
  {\bibfnamefont {E.}~\bibnamefont {Lhotel}}, \ and\ \bibinfo {author}
  {\bibfnamefont {C.}~\bibnamefont {Paulsen}},\ }\href
  {http://dx.doi.org/10.1038/35098048} {\bibfield  {journal} {\bibinfo
  {journal} {Nature}\ }\textbf {\bibinfo {volume} {413}},\ \bibinfo {pages}
  {613} (\bibinfo {year} {2001})}\BibitemShut {NoStop}%
\bibitem [{\citenamefont {Huy}\ \emph {et~al.}(2007)\citenamefont {Huy},
  \citenamefont {Gasparini}, \citenamefont {de~Nijs}, \citenamefont {Huang},
  \citenamefont {Klaasse}, \citenamefont {Gortenmulder}, \citenamefont
  {de~Visser}, \citenamefont {Hamann}, \citenamefont {G\"orlach},\ and\
  \citenamefont {L\"ohneysen}}]{Huy2007}%
  \BibitemOpen
  \bibfield  {author} {\bibinfo {author} {\bibfnamefont {N.~T.}\ \bibnamefont
  {Huy}}, \bibinfo {author} {\bibfnamefont {A.}~\bibnamefont {Gasparini}},
  \bibinfo {author} {\bibfnamefont {D.~E.}\ \bibnamefont {de~Nijs}}, \bibinfo
  {author} {\bibfnamefont {Y.}~\bibnamefont {Huang}}, \bibinfo {author}
  {\bibfnamefont {J.~C.~P.}\ \bibnamefont {Klaasse}}, \bibinfo {author}
  {\bibfnamefont {T.}~\bibnamefont {Gortenmulder}}, \bibinfo {author}
  {\bibfnamefont {A.}~\bibnamefont {de~Visser}}, \bibinfo {author}
  {\bibfnamefont {A.}~\bibnamefont {Hamann}}, \bibinfo {author} {\bibfnamefont
  {T.}~\bibnamefont {G\"orlach}}, \ and\ \bibinfo {author} {\bibfnamefont
  {H.~v.}\ \bibnamefont {L\"ohneysen}},\ }\href {\doibase
  10.1103/PhysRevLett.99.067006} {\bibfield  {journal} {\bibinfo  {journal}
  {Phys. Rev. Lett.}\ }\textbf {\bibinfo {volume} {99}},\ \bibinfo {pages}
  {067006} (\bibinfo {year} {2007})}\BibitemShut {NoStop}%
\bibitem [{\citenamefont {Sato}\ and\ \citenamefont {Ando}(2017)}]{Sato2017}%
  \BibitemOpen
  \bibfield  {author} {\bibinfo {author} {\bibfnamefont {M.}~\bibnamefont
  {Sato}}\ and\ \bibinfo {author} {\bibfnamefont {Y.}~\bibnamefont {Ando}},\
  }\href {http://stacks.iop.org/0034-4885/80/i=7/a=076501} {\bibfield
  {journal} {\bibinfo  {journal} {Reports on Progress in Physics}\ }\textbf
  {\bibinfo {volume} {80}},\ \bibinfo {pages} {076501} (\bibinfo {year}
  {2017})}\BibitemShut {NoStop}%
\bibitem [{\citenamefont {Aoki}\ \emph
  {et~al.}(2019{\natexlab{a}})\citenamefont {Aoki}, \citenamefont {Nakamura},
  \citenamefont {Honda}, \citenamefont {Li}, \citenamefont {Homma},
  \citenamefont {Shimizu}, \citenamefont {Sato}, \citenamefont {Knebel},
  \citenamefont {Brison}, \citenamefont {Pourret}, \citenamefont {Braithwaite},
  \citenamefont {Lapertot}, \citenamefont {Niu}, \citenamefont {Vališka},
  \citenamefont {Harima},\ and\ \citenamefont {Flouquet}}]{Aoki2019}%
  \BibitemOpen
  \bibfield  {author} {\bibinfo {author} {\bibfnamefont {D.}~\bibnamefont
  {Aoki}}, \bibinfo {author} {\bibfnamefont {A.}~\bibnamefont {Nakamura}},
  \bibinfo {author} {\bibfnamefont {F.}~\bibnamefont {Honda}}, \bibinfo
  {author} {\bibfnamefont {D.}~\bibnamefont {Li}}, \bibinfo {author}
  {\bibfnamefont {Y.}~\bibnamefont {Homma}}, \bibinfo {author} {\bibfnamefont
  {Y.}~\bibnamefont {Shimizu}}, \bibinfo {author} {\bibfnamefont {Y.~J.}\
  \bibnamefont {Sato}}, \bibinfo {author} {\bibfnamefont {G.}~\bibnamefont
  {Knebel}}, \bibinfo {author} {\bibfnamefont {J.-P.}\ \bibnamefont {Brison}},
  \bibinfo {author} {\bibfnamefont {A.}~\bibnamefont {Pourret}}, \bibinfo
  {author} {\bibfnamefont {D.}~\bibnamefont {Braithwaite}}, \bibinfo {author}
  {\bibfnamefont {G.}~\bibnamefont {Lapertot}}, \bibinfo {author}
  {\bibfnamefont {Q.}~\bibnamefont {Niu}}, \bibinfo {author} {\bibfnamefont
  {M.}~\bibnamefont {Vališka}}, \bibinfo {author} {\bibfnamefont
  {H.}~\bibnamefont {Harima}}, \ and\ \bibinfo {author} {\bibfnamefont
  {J.}~\bibnamefont {Flouquet}},\ }\href {\doibase 10.7566/jpsj.88.043702}
  {\bibfield  {journal} {\bibinfo  {journal} {J. Phys. Soc. Jpn.}\ }\textbf
  {\bibinfo {volume} {88}},\ \bibinfo {pages} {043702} (\bibinfo {year}
  {2019}{\natexlab{a}})}\BibitemShut {NoStop}%
\bibitem [{\citenamefont {Knebel}\ \emph {et~al.}(2019)\citenamefont {Knebel},
  \citenamefont {Knafo}, \citenamefont {Pourret}, \citenamefont {Niu},
  \citenamefont {Vališka}, \citenamefont {Braithwaite}, \citenamefont
  {Lapertot}, \citenamefont {Nardone}, \citenamefont {Zitouni}, \citenamefont
  {Mishra}, \citenamefont {Sheikin}, \citenamefont {Seyfarth}, \citenamefont
  {Brison}, \citenamefont {Aoki},\ and\ \citenamefont {Flouquet}}]{Knebel2019}%
  \BibitemOpen
  \bibfield  {author} {\bibinfo {author} {\bibfnamefont {G.}~\bibnamefont
  {Knebel}}, \bibinfo {author} {\bibfnamefont {W.}~\bibnamefont {Knafo}},
  \bibinfo {author} {\bibfnamefont {A.}~\bibnamefont {Pourret}}, \bibinfo
  {author} {\bibfnamefont {Q.}~\bibnamefont {Niu}}, \bibinfo {author}
  {\bibfnamefont {M.}~\bibnamefont {Vališka}}, \bibinfo {author}
  {\bibfnamefont {D.}~\bibnamefont {Braithwaite}}, \bibinfo {author}
  {\bibfnamefont {G.}~\bibnamefont {Lapertot}}, \bibinfo {author}
  {\bibfnamefont {M.}~\bibnamefont {Nardone}}, \bibinfo {author} {\bibfnamefont
  {A.}~\bibnamefont {Zitouni}}, \bibinfo {author} {\bibfnamefont
  {S.}~\bibnamefont {Mishra}}, \bibinfo {author} {\bibfnamefont
  {I.}~\bibnamefont {Sheikin}}, \bibinfo {author} {\bibfnamefont
  {G.}~\bibnamefont {Seyfarth}}, \bibinfo {author} {\bibfnamefont {J.-P.}\
  \bibnamefont {Brison}}, \bibinfo {author} {\bibfnamefont {D.}~\bibnamefont
  {Aoki}}, \ and\ \bibinfo {author} {\bibfnamefont {J.}~\bibnamefont
  {Flouquet}},\ }\href {\doibase 10.7566/jpsj.88.063707} {\bibfield  {journal}
  {\bibinfo  {journal} {J. Phys. Soc. Jpn.}\ }\textbf {\bibinfo {volume}
  {88}},\ \bibinfo {pages} {063707} (\bibinfo {year} {2019})}\BibitemShut
  {NoStop}%
\bibitem [{\citenamefont {Metz}\ \emph {et~al.}(2019)\citenamefont {Metz},
  \citenamefont {Bae}, \citenamefont {Ran}, \citenamefont {Liu}, \citenamefont
  {Eo}, \citenamefont {Fuhrman}, \citenamefont {Agterberg}, \citenamefont
  {Anlage}, \citenamefont {Butch},\ and\ \citenamefont {Paglione}}]{Metz2019}%
  \BibitemOpen
  \bibfield  {author} {\bibinfo {author} {\bibfnamefont {T.}~\bibnamefont
  {Metz}}, \bibinfo {author} {\bibfnamefont {S.}~\bibnamefont {Bae}}, \bibinfo
  {author} {\bibfnamefont {S.}~\bibnamefont {Ran}}, \bibinfo {author}
  {\bibfnamefont {I.-L.}\ \bibnamefont {Liu}}, \bibinfo {author} {\bibfnamefont
  {Y.~S.}\ \bibnamefont {Eo}}, \bibinfo {author} {\bibfnamefont {W.~T.}\
  \bibnamefont {Fuhrman}}, \bibinfo {author} {\bibfnamefont {D.~F.}\
  \bibnamefont {Agterberg}}, \bibinfo {author} {\bibfnamefont {S.}~\bibnamefont
  {Anlage}}, \bibinfo {author} {\bibfnamefont {N.~P.}\ \bibnamefont {Butch}}, \
  and\ \bibinfo {author} {\bibfnamefont {J.}~\bibnamefont {Paglione}},\
  }\href@noop {} {\bibfield  {journal} {\bibinfo  {journal} {arXiv}\ }
  (\bibinfo {year} {2019})}\BibitemShut {NoStop}%
\bibitem [{\citenamefont {Jiao}\ \emph {et~al.}(2019)\citenamefont {Jiao},
  \citenamefont {Wang}, \citenamefont {Ran}, \citenamefont {Rodriguez},
  \citenamefont {Sigrist}, \citenamefont {Wang}, \citenamefont {Butch},\ and\
  \citenamefont {Madhavan}}]{Jiao2019}%
  \BibitemOpen
  \bibfield  {author} {\bibinfo {author} {\bibfnamefont {L.}~\bibnamefont
  {Jiao}}, \bibinfo {author} {\bibfnamefont {Z.}~\bibnamefont {Wang}}, \bibinfo
  {author} {\bibfnamefont {S.}~\bibnamefont {Ran}}, \bibinfo {author}
  {\bibfnamefont {J.~O.}\ \bibnamefont {Rodriguez}}, \bibinfo {author}
  {\bibfnamefont {M.}~\bibnamefont {Sigrist}}, \bibinfo {author} {\bibfnamefont
  {Z.}~\bibnamefont {Wang}}, \bibinfo {author} {\bibfnamefont {N.}~\bibnamefont
  {Butch}}, \ and\ \bibinfo {author} {\bibfnamefont {V.}~\bibnamefont
  {Madhavan}},\ }\href@noop {} {\bibfield  {journal} {\bibinfo  {journal}
  {arXiv}\ } (\bibinfo {year} {2019})}\BibitemShut {NoStop}%
\bibitem [{\citenamefont {Hutanu}\ \emph {et~al.}(2019)\citenamefont {Hutanu},
  \citenamefont {Deng}, \citenamefont {Ran}, \citenamefont {Fuhrman},
  \citenamefont {Thoma},\ and\ \citenamefont {Butch}}]{Hutanu2019}%
  \BibitemOpen
  \bibfield  {author} {\bibinfo {author} {\bibfnamefont {V.}~\bibnamefont
  {Hutanu}}, \bibinfo {author} {\bibfnamefont {H.}~\bibnamefont {Deng}},
  \bibinfo {author} {\bibfnamefont {S.}~\bibnamefont {Ran}}, \bibinfo {author}
  {\bibfnamefont {W.~T.}\ \bibnamefont {Fuhrman}}, \bibinfo {author}
  {\bibfnamefont {H.}~\bibnamefont {Thoma}}, \ and\ \bibinfo {author}
  {\bibfnamefont {N.~P.}\ \bibnamefont {Butch}},\ }\href@noop {} {\bibfield
  {journal} {\bibinfo  {journal} {arXiv}\ } (\bibinfo {year}
  {2019})}\BibitemShut {NoStop}%
\bibitem [{\citenamefont {Batlogg}\ \emph {et~al.}(1987)\citenamefont
  {Batlogg}, \citenamefont {Bishop}, \citenamefont {Bucher}, \citenamefont
  {Golding}, \citenamefont {Ramirez}, \citenamefont {Fisk}, \citenamefont
  {Smith},\ and\ \citenamefont {Ott}}]{Batlogg1987}%
  \BibitemOpen
  \bibfield  {author} {\bibinfo {author} {\bibfnamefont {B.}~\bibnamefont
  {Batlogg}}, \bibinfo {author} {\bibfnamefont {D.~J.}\ \bibnamefont {Bishop}},
  \bibinfo {author} {\bibfnamefont {E.}~\bibnamefont {Bucher}}, \bibinfo
  {author} {\bibfnamefont {B.}~\bibnamefont {Golding}}, \bibinfo {author}
  {\bibfnamefont {A.~P.}\ \bibnamefont {Ramirez}}, \bibinfo {author}
  {\bibfnamefont {Z.}~\bibnamefont {Fisk}}, \bibinfo {author} {\bibfnamefont
  {J.~L.}\ \bibnamefont {Smith}}, \ and\ \bibinfo {author} {\bibfnamefont
  {H.~R.}\ \bibnamefont {Ott}},\ }\href
  {http://www.sciencedirect.com/science/article/pii/0304885387906329}
  {\bibfield  {journal} {\bibinfo  {journal} {Journal of Magnetism and Magnetic
  Materials}\ }\textbf {\bibinfo {volume} {63-64}},\ \bibinfo {pages} {441}
  (\bibinfo {year} {1987})}\BibitemShut {NoStop}%
\bibitem [{\citenamefont {Brison}\ \emph {et~al.}(1989)\citenamefont {Brison},
  \citenamefont {Laborde}, \citenamefont {Jaccard}, \citenamefont {Flouquet},
  \citenamefont {Morin}, \citenamefont {Fisk},\ and\ \citenamefont
  {Smith}}]{Brison1989}%
  \BibitemOpen
  \bibfield  {author} {\bibinfo {author} {\bibfnamefont {J.~P.}\ \bibnamefont
  {Brison}}, \bibinfo {author} {\bibfnamefont {O.}~\bibnamefont {Laborde}},
  \bibinfo {author} {\bibfnamefont {D.}~\bibnamefont {Jaccard}}, \bibinfo
  {author} {\bibfnamefont {J.}~\bibnamefont {Flouquet}}, \bibinfo {author}
  {\bibfnamefont {P.}~\bibnamefont {Morin}}, \bibinfo {author} {\bibfnamefont
  {Z.}~\bibnamefont {Fisk}}, \ and\ \bibinfo {author} {\bibfnamefont {J.~L.}\
  \bibnamefont {Smith}},\ }\href
  {https://doi.org/10.1051/jphys:0198900500180279500} {\bibfield  {journal}
  {\bibinfo  {journal} {J. Phys. France}\ }\textbf {\bibinfo {volume} {50}},\
  \bibinfo {pages} {2795} (\bibinfo {year} {1989})}\BibitemShut {NoStop}%
\bibitem [{\citenamefont {Park}\ \emph {et~al.}(2008)\citenamefont {Park},
  \citenamefont {Sidorov}, \citenamefont {Ronning}, \citenamefont {Zhu},
  \citenamefont {Tokiwa}, \citenamefont {Lee}, \citenamefont {Bauer},
  \citenamefont {Movshovich}, \citenamefont {Sarrao},\ and\ \citenamefont
  {Thompson}}]{Park2008}%
  \BibitemOpen
  \bibfield  {author} {\bibinfo {author} {\bibfnamefont {T.}~\bibnamefont
  {Park}}, \bibinfo {author} {\bibfnamefont {V.~A.}\ \bibnamefont {Sidorov}},
  \bibinfo {author} {\bibfnamefont {F.}~\bibnamefont {Ronning}}, \bibinfo
  {author} {\bibfnamefont {J.-X.}\ \bibnamefont {Zhu}}, \bibinfo {author}
  {\bibfnamefont {Y.}~\bibnamefont {Tokiwa}}, \bibinfo {author} {\bibfnamefont
  {H.}~\bibnamefont {Lee}}, \bibinfo {author} {\bibfnamefont {E.~D.}\
  \bibnamefont {Bauer}}, \bibinfo {author} {\bibfnamefont {R.}~\bibnamefont
  {Movshovich}}, \bibinfo {author} {\bibfnamefont {J.~L.}\ \bibnamefont
  {Sarrao}}, \ and\ \bibinfo {author} {\bibfnamefont {J.~D.}\ \bibnamefont
  {Thompson}},\ }\href {https://doi.org/10.1038/nature07431} {\bibfield
  {journal} {\bibinfo  {journal} {Nature}\ }\textbf {\bibinfo {volume} {456}},\
  \bibinfo {pages} {366} (\bibinfo {year} {2008})}\BibitemShut {NoStop}%
\bibitem [{\citenamefont {Brando}\ \emph {et~al.}(2016)\citenamefont {Brando},
  \citenamefont {Belitz}, \citenamefont {Grosche},\ and\ \citenamefont
  {Kirkpatrick}}]{Brando2016}%
  \BibitemOpen
  \bibfield  {author} {\bibinfo {author} {\bibfnamefont {M.}~\bibnamefont
  {Brando}}, \bibinfo {author} {\bibfnamefont {D.}~\bibnamefont {Belitz}},
  \bibinfo {author} {\bibfnamefont {F.~M.}\ \bibnamefont {Grosche}}, \ and\
  \bibinfo {author} {\bibfnamefont {T.~R.}\ \bibnamefont {Kirkpatrick}},\
  }\href {https://link.aps.org/doi/10.1103/RevModPhys.88.025006} {\bibfield
  {journal} {\bibinfo  {journal} {RMP}\ }\textbf {\bibinfo {volume} {88}},\
  \bibinfo {pages} {025006} (\bibinfo {year} {2016})}\BibitemShut {NoStop}%
\bibitem [{\citenamefont {Schmakat}\ \emph {et~al.}(2015)\citenamefont
  {Schmakat}, \citenamefont {Wagner}, \citenamefont {Ritz}, \citenamefont
  {Bauer}, \citenamefont {Brando}, \citenamefont {Deppe}, \citenamefont
  {Duncan}, \citenamefont {Duvinage}, \citenamefont {Franz}, \citenamefont
  {Geibel}, \citenamefont {Grosche}, \citenamefont {Hirschberger},
  \citenamefont {Hradil}, \citenamefont {Meven}, \citenamefont {Neubauer},
  \citenamefont {Schulz}, \citenamefont {Senyshyn}, \citenamefont {Süllow},
  \citenamefont {Pedersen}, \citenamefont {Böni},\ and\ \citenamefont
  {Pfleiderer}}]{Schmakat2015}%
  \BibitemOpen
  \bibfield  {author} {\bibinfo {author} {\bibfnamefont {P.}~\bibnamefont
  {Schmakat}}, \bibinfo {author} {\bibfnamefont {M.}~\bibnamefont {Wagner}},
  \bibinfo {author} {\bibfnamefont {R.}~\bibnamefont {Ritz}}, \bibinfo {author}
  {\bibfnamefont {A.}~\bibnamefont {Bauer}}, \bibinfo {author} {\bibfnamefont
  {M.}~\bibnamefont {Brando}}, \bibinfo {author} {\bibfnamefont
  {M.}~\bibnamefont {Deppe}}, \bibinfo {author} {\bibfnamefont
  {W.}~\bibnamefont {Duncan}}, \bibinfo {author} {\bibfnamefont
  {C.}~\bibnamefont {Duvinage}}, \bibinfo {author} {\bibfnamefont
  {C.}~\bibnamefont {Franz}}, \bibinfo {author} {\bibfnamefont
  {C.}~\bibnamefont {Geibel}}, \bibinfo {author} {\bibfnamefont {F.~M.}\
  \bibnamefont {Grosche}}, \bibinfo {author} {\bibfnamefont {M.}~\bibnamefont
  {Hirschberger}}, \bibinfo {author} {\bibfnamefont {K.}~\bibnamefont
  {Hradil}}, \bibinfo {author} {\bibfnamefont {M.}~\bibnamefont {Meven}},
  \bibinfo {author} {\bibfnamefont {A.}~\bibnamefont {Neubauer}}, \bibinfo
  {author} {\bibfnamefont {M.}~\bibnamefont {Schulz}}, \bibinfo {author}
  {\bibfnamefont {A.}~\bibnamefont {Senyshyn}}, \bibinfo {author}
  {\bibfnamefont {S.}~\bibnamefont {Süllow}}, \bibinfo {author} {\bibfnamefont
  {B.}~\bibnamefont {Pedersen}}, \bibinfo {author} {\bibfnamefont
  {P.}~\bibnamefont {Böni}}, \ and\ \bibinfo {author} {\bibfnamefont
  {C.}~\bibnamefont {Pfleiderer}},\ }\href
  {https://doi.org/10.1140/epjst/e2015-02445-4} {\bibfield  {journal} {\bibinfo
   {journal} {The European Physical Journal Special Topics}\ }\textbf {\bibinfo
  {volume} {224}},\ \bibinfo {pages} {1041} (\bibinfo {year}
  {2015})}\BibitemShut {NoStop}%
\bibitem [{\citenamefont {Si}(2006)}]{Si2006}%
  \BibitemOpen
  \bibfield  {author} {\bibinfo {author} {\bibfnamefont {Q.}~\bibnamefont
  {Si}},\ }\href
  {http://www.sciencedirect.com/science/article/pii/S092145260600007X}
  {\bibfield  {journal} {\bibinfo  {journal} {Physica B: Condensed Matter}\
  }\textbf {\bibinfo {volume} {378-380}},\ \bibinfo {pages} {23} (\bibinfo
  {year} {2006})}\BibitemShut {NoStop}%
\bibitem [{\citenamefont {Si}(2010)}]{Si2010}%
  \BibitemOpen
  \bibfield  {author} {\bibinfo {author} {\bibfnamefont {Q.}~\bibnamefont
  {Si}},\ }\href {\doibase 10.1002/pssb.200983082} {\bibfield  {journal}
  {\bibinfo  {journal} {physica status solidi (b)}\ }\textbf {\bibinfo {volume}
  {247}},\ \bibinfo {pages} {476} (\bibinfo {year} {2010})}\BibitemShut
  {NoStop}%
\bibitem [{\citenamefont {Aoki}\ \emph
  {et~al.}(2019{\natexlab{b}})\citenamefont {Aoki}, \citenamefont {Ishida},\
  and\ \citenamefont {Flouquet}}]{Aoki2019a}%
  \BibitemOpen
  \bibfield  {author} {\bibinfo {author} {\bibfnamefont {D.}~\bibnamefont
  {Aoki}}, \bibinfo {author} {\bibfnamefont {K.}~\bibnamefont {Ishida}}, \ and\
  \bibinfo {author} {\bibfnamefont {J.}~\bibnamefont {Flouquet}},\ }\href
  {\doibase 10.7566/jpsj.88.022001} {\bibfield  {journal} {\bibinfo  {journal}
  {J. Phys. Soc. Jpn.}\ }\textbf {\bibinfo {volume} {88}},\ \bibinfo {pages}
  {022001} (\bibinfo {year} {2019}{\natexlab{b}})}\BibitemShut {NoStop}%
\bibitem [{\citenamefont {Mineev}(2017)}]{Mineev2017a}%
  \BibitemOpen
  \bibfield  {author} {\bibinfo {author} {\bibfnamefont {V.~P.}\ \bibnamefont
  {Mineev}},\ }\href {http://stacks.iop.org/1063-7869/60/i=2/a=121} {\bibfield
  {journal} {\bibinfo  {journal} {Physics-Uspekhi}\ }\textbf {\bibinfo {volume}
  {60}},\ \bibinfo {pages} {121} (\bibinfo {year} {2017})}\BibitemShut
  {NoStop}%
\bibitem [{\citenamefont {Smith}\ and\ \citenamefont {Chu}(1967)}]{Smith1967}%
  \BibitemOpen
  \bibfield  {author} {\bibinfo {author} {\bibfnamefont {T.~F.}\ \bibnamefont
  {Smith}}\ and\ \bibinfo {author} {\bibfnamefont {C.~W.}\ \bibnamefont
  {Chu}},\ }\href {https://link.aps.org/doi/10.1103/PhysRev.159.353} {\bibfield
   {journal} {\bibinfo  {journal} {PR}\ }\textbf {\bibinfo {volume} {159}},\
  \bibinfo {pages} {353} (\bibinfo {year} {1967})}\BibitemShut {NoStop}%
\end{thebibliography}%

\clearpage

\end{document}